# Five Modular Redundancy with Mitigation Technique to Recover the Error Module

Haryono [1], Jazi Eko Istiyanto, Agus Harjoko, Agfianto Eko Putra
Doctorate Program in Computer Science, Department of Computer Science and Electronics
Faculty of Mathematics and Natural Sciences, Gadjah Mada University
Yogyakarta, Indonesia

*Abstract*— Hazard radiation can lead the system fault therefore Fault Tolerance is required. Fault Tolerant is a system, which is designed to keep operations running, despite the degradation in the specific module is happening. Many fault tolerances have been developed to handle the problem, to find the most robust and efficient in the possible technology. This paper will present the Five Modular Redundancy (FMR) with Mitigation Technique to Recover the Error Module. With Dynamic Partial Reconfiguration technology that have already available today, such fault tolerance technique can be implemented successfully. The project showed the robustness of the system is increased and module which is error can be recovered immediately.

## I. Introduction

Fault is a changed in the value of a variable or unexpected logic in the system hardware, failure is the inability of a system to perform the operation from predefined requirements [1]. A system fault has a chance for failure, it requires a Fault Tolerant system. Fault Tolerant is a character system that is designed to continue to run its operations despite the degradation of function in the specific module, do not stop completely when the failure occurred on a particular module [2].

Fault tolerance design in [3] and [4] is using Triple Modular Redundancy (TMR) by means triplicate a module or a particular function. In TMR at least two modules produce the same results, then the system is considered to be running correct. Since in the orbit in such area is having many radiations, as quoted in [5] the TMR design is not enough to mitigate the entire fault that is occurs, it may occur at two memories at the same time and same position and then give two modules in error result. In [5] Nine Modular Redundancy (NMR) has been developed to try to handle the TMR problem, but leads in using a lot of resources. To full fill the gap between those two designs we therefore create a new methodology, we called Five Modular Redundancy (FMR) with Mitigation Technique to Recover the Error Module. In [5] use nine redundancies, but have not been implemented a recover technique when some module is an error, therefore degradation of a system cannot be avoided. By implementing recovering technique to the error module such degradation is kept as minimal as possible. To overcome in TMR technique, this design will handle the problem about radiation bombardment that makes two error modules at the same time.

The scope of this project is about FMR design with mitigation technique using DPR technology, we assume internal design is free from a fault in which this Fault Tolerance aims to handle a fault that is caused by external factors e.g. hazard radiations. This paper will show the design, implementation and testing of Five Modular Redundancy with mitigation technique to recover the error module. The testing showed, the design that is developed able to handle the error that happen in the two modules at the same time and able to detect and recover the error module.

## II. Design of Five Modular Redundancy with Mitigation Technique to Recover the Error Module

Five Modular Redundancy (FMR) is the technique to duplicate the same module to five times. The system still has correct resulted since the modules are having correct result with at least three modules, when the error is happening in the two modules the system can still tolerance to such error. Using Dynamic Partial Reconfiguration (DPR) will not interrupt the system which is running, therefore mitigation to the error module can be done without disturbing the system. DPR can be done very fast and low power consumption because we only reconfigure to the partial area of an FPGA.

Image 1 shows the design of the FMR, having five modules that are identical, the output of each module will be sent to Voter and Error Detector, then the Voter will vote the result from each module and find the output. Error Detector is the important role, it plays to detect the modules which are error, one the error is detected, Error Detector will send the data to Dynamic Partial Reconfiguration (DPR) system. DPR will then do the reconfiguration to the module which is an error.

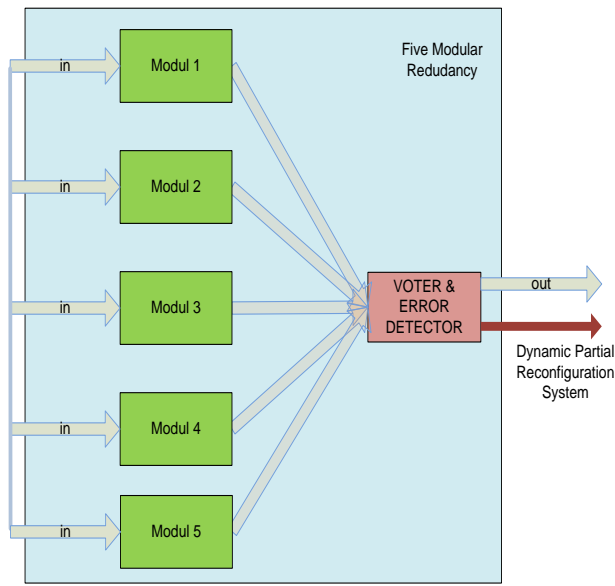

Image 1. Five Modular Redundancy Design

Voter architecture can be seen in the Image 2. At least three modules which are correct, the FMR will have a correct result. The output of each module will be sent to the "*And*" logic as shown in the Image 2 (a) with a voter logic combination that is shown in the Image 2(b).

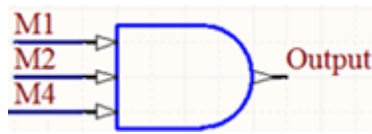

(a)

$F =$ (M1 and M2 and M3 ) or
(M1 and M2 and M4 ) or
(M1 and M2 and M5 ) or
(M1 and M3 and M4 ) or
(M1 and M3 and M5 ) or
(M1 and M4 and M5 ) or
(M2 and M3 and M4 ) or
(M2 and M3 and M5 ) or
(M2 and M4 and M5 ) or
(M3 and M4 and M5 );

(b)

Image 2. Voter of FMR

The result of voter will be compared to each module, the modules that have output not same with the result of voter will be the fault/error module. The error Detector variable holds the modules which are error and which are correct, it is an array variable that has length five bits. Following is a pseudo code configuration of error detector:

*if (F is equal M1) then ErrorDetectorVariable[0] =
1 else ErrorDetectorVariable[0] = 0
if (F is equal M2) then ErrorDetectorVariable[1] =
1 else ErrorDetectorVariable[1] = 0
if (F is equal M3) then ErrorDetectorVariable[2] =
1 else ErrorDetectorVariable[2] = 0
if (F is equal M4) then ErrorDetectorVariable[3] =
1 else ErrorDetectorVariable[3] = 0
if (F is equal M5) then ErrorDetectorVariable[4] =
1 else ErrorDetectorVariable[4] = 0*

The structure of DPR is shown in the Image 3. Microblaze is a microprocessor that handles the Dynamic Partial Configuration process. The Bit stream is saved in the non volatile memory, the frame data contain FPGA Location, Configuration data and Check Sum. Error Detector will send the data when the error is detected trough bus to Microblaze processor. When Microblaze receives the data from Error Detector, the Microblaze read the data of partial Bitstream from memory compact flash, then the data is sent trough ICAP, ICAP will do the configuration by placing the configuration data to the portion of the FPGA.

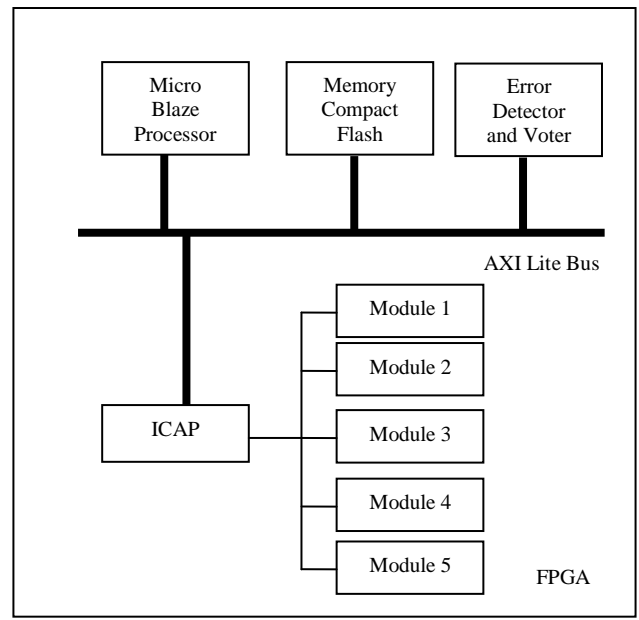

Image 3. Structure of DPR

III. IMPLEMENTATION

Recovering modules which are error is the important part in this design because we expect the system will have a better reliability, once it is detected the error in a module the system should recover the module. Recover step can be split into the following stage: **first** is checking the output of each module in the Detector Unit, **second** is when error module is detected interrupt to the DPR system, **third** is DPR system will receive new interrupt to recover a module which is an error and then doing recovering by dynamic partial reconfiguration correspond to the error module.

Checking the output of each module in the Detector Unit can be done by putting the VHDL checking in the Voter Unit. The VHDL code is shown in the following code:

```
ErrorDetector(0) < '0' when FtResult = PR_Input1 else '1';
ErrorDetector(1) < '0' when FtResult = PR_Input2 else '1';
ErrorDetector(2) <='0' when FtResult = PR_Input3 else '1';
ErrorDetector(3) < '0' when FtResult = PR_Input4 else '1';
ErrorDetector(4) <='0' when FtResult = PR_Input5 else '1';
```

It is checking the result of the FMR Fault Tolerance output, if the output of particular module is not same with the output of the FMR Fault Tolerance output then we recognize it is an error module. After the error is detected, we do an action to recover the error module. In Microblaze processor we check the *ErrorDetector* data regularly, we can adjust the interval to check the *ErrorDetector* depends on the need. By reading the address of the custom Intellectual Property (IP) of FMR fault tolerance to particular *ErrorDetector* variable, we can get the information which one the module that is error, following is the code to get the data:

```
int error_detector_module =
Xil_In32(XPAR_DUALMODEFT_0_BASEADDR);
int *bits = get_bits(result, 5);
if(bits[0] == 1 )
{
        PR_Action('1'); // Action for DPR
}
```

When the *error_detector_module* is more than zero that means the error in module is detected, the error module can be checked by the position of bit, Least Significant Bit (LSB) will be the first module followed by second module and so forth. In the above code shown that *error_detector_module* is converted to array so that can be easy to check the error module.

DPR flow is described in the Image 4, When the error module is recognized, the Microblaze processor will read the bit stream in the compact flash non volatile memory, if the error module is first module then read "module1.bit" file, if the second module then reads "module2.bit" file and so forth. Reading trough System ACE and save to the system ACE Buffer. Then loop to each data and send the data to ICAP, there is *XHwIcap_DeviceWrite* function to write the data to ICAP. ICAP will handle of the rest, to where the reconfiguration data will be place to the Memory location of FPGA.

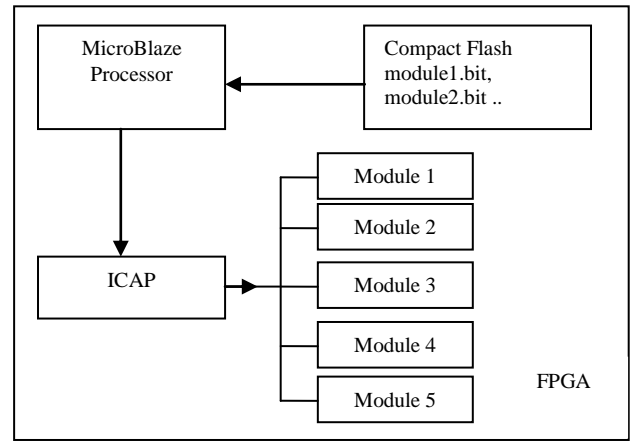

Image 4. DPR data flow to recover error module

IV. TESTING AND DISCUSSION

The design which is offered is to overcome about the two models of fault tolerances that have been developed, they are TMR with mitigation technique and NMR. Table 1 is the comparison between them and the design that is offered. TMR with mitigation technique will end when the two modules is an error at the same time, this is can be happened in the space, to handle this so we implemented FMR, in which limit tolerance to a fault can be up to two modules. NMR that has been developed can still work if there are three error modules at the same time, but in the mean time during operation the module that is error cannot be recovered from erroneous, this will make the system become degraded in several times. To overcome the NMR we implement mitigation technique that can mitigate or recover the error modules.

TABLE I. COMPARISON BETWEEN FAULT TOLERANCE TECHNIQUES

| Technique | Limit Tolerance To Fault | Mitigation to Error module |
|---|---|---|
| TMR with mitigation technique | Up to one error module | Yes |
| NMR with TMR scheme | Up to three error modules | No |
| FMR with mitigation technique | Up to two error modules | Yes |

Testing of fault tolerance can be done by providing a direct test on the hardware of FPGA device, by giving a large ion injection or given power supply disturbance [6]. However, this method is relatively expensive and difficult to obtain expected environment. Another method is using partial reconfiguration, [7] demonstrated that the partial reconfiguration is an effective way to perform fault injection. This method is also done by [8] and the same thing is done by [9], so authors chose the second way. The structure of the test is not different with the structure of DPR that was described in the image 3, in the testing, we add the computer connecting to the MicroBlaze to acquire the data during a testing that is conducted, it is shown in the image 5.

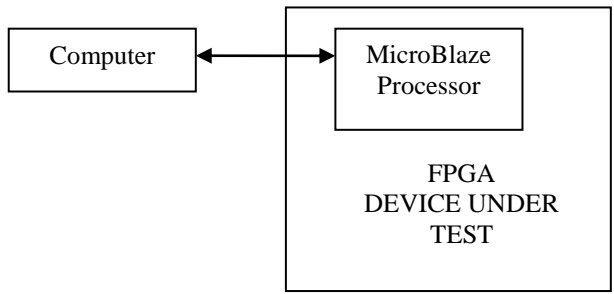

Image 5. Structure of testing based on fault injector using partial reconfiguration.

The original partial bit stream is made corrupt or blank. Fault injection with partial reconfiguration, by changing directly to a partial Bitstream files are at risk because can permanently damage the FPGA device, in which the bits that have been extracted from the tools provided by Xilinx is like a program "exe" that is gotten from compiling windows programs. There is *Jbits* program that can act to change the partial bits, then dynamically reconfigured the FPGA. Currently we choose by making a blank configuration of the module that is tested, this method can be done trough partial reconfiguration using the Microblaze processor.

We send a command from computer to Microblaze to do the reconfiguration with blank reconfiguration to a particular module. A module which is applied to reconfigure by blank module is selected randomly each 500 ms, and then we analyze the data after reconfiguration with blank module. The result is following: first is after various randomize reconfiguration to a particular module, mitigation or recovering to a particular module that is blank can be reconfigured again to become a correct module without giving effect to a system that is running. Second is FMR fault tolerance still works as expected by sending some data to a module and then a module give feedback as expected. We also made two modules at the same time to be in error state, the FMR also can have a correct result.

The speed to recover the module which is error depends on the size of the bit stream. To get the speed we apply calculation in real code application. We add the AXI Timer IP to our DPR system. We initiated and started it by following code:

*int Status = XTmrCtr_Initialize(&TimerCounterInst, XPAR_AXI_TIMER_0_DEVICE_ID);*
*XTmrCtr_Start(&TimerCounterInst,0);*

The methodology to take the time which is required to do recovery to each module is following: first is resetting the timer counter because if we don't reset to some period the counter will overflow. The second is getting the counter register variable. Then do the recovery. After finishing in recovery then we take the counter register again and put it in the variable. Following is the code that we have implemented:

*XTmrCtr_Reset(&TimerCounterInst,0);*
*int startExe =*
*XTmrCtr_GetTimerCounterReg(XPAR_AXI_TIMER_0_BASEADDR,XPAR_AXI_TIMER_0_DEVICE_ID);*
**get_modul_error_and_recovery();**
*int endExe =*
*XTmrCtr_GetTimerCounterReg(XPAR_AXI_TIMER_0_BASEADDR,XPAR_AXI_TIMER_0_DEVICE_ID);*

Our Microblaze processor speed is 100MHz, one integer represents one clock cycle, in which one clock cycle is 10 ns. Table II shows the speed of recovering to each module, it includes reading the file in Compact flash and Writing to the ICAP. The size of the module is varied depends on the number of resources which is used, it can be different because when we use *Pblock function* to draw to a *device* in Plan Ahead is varied.

TABLE II.    SPEED OF RECOVERING ERROR MODULE

| Module | Size (KB) | Speed (ms) |
|---|---|---|
| 1 | 128 | 224.93 |
| 2 | 120 | 209.66 |
| 3 | 81 | 141.59 |
| 4 | 128 | 225.00 |
| 5 | 142 | 261.57 |

We need to estimate how much power that is required if using FMR. The calculation uses *Xilinx Power Estimator* (XPE) for Virtex 6. Table 3 is a calculation of power that is required when each module is using one microblaze processor. There are 1573 LUT Logic, 103 Distributed RAM and 1456 flip flop. In that figure showed, the power that is required for each module is 0.010 w. Therefore, if using FMR we just multiply by 5 so become 0.050 w. If using FMR we multiply by 3 become 0.030 w. By using FMR we must consider about the power budget, will it satisfy with the budget or not. By using XPE to do a calculation estimation of power consumption above, we can decide whether the fault tolerance can be implemented.

TABLE III. ESTIMATION FOR EACH MODULE USING XPE

| Name | Clock (MHz) | LUTs as Logic | LUTs as Shift Registers | LUTs as Distributed RAMs | FFs | Toggle Rate | Average Fanout | Signal Rate (Mtr/s) | Power (W) |
|---|---|---|---|---|---|---|---|---|---|
| module1 | 100.0 | 1456 | 0 | 103 | 1573 | 12.5% | 3.00 | 12.5 | 0.010 |
| module2 | 100.0 | 1456 | 0 | 103 | 1573 | 12.5% | 3.00 | 12.5 | 0.010 |
| module3 | 100.0 | 1456 | 0 | 103 | 1573 | 12.5% | 3.00 | 12.5 | 0.010 |
| module4 | 100.0 | 1456 | 0 | 103 | 1573 | 12.5% | 3.00 | 12.5 | 0.010 |
| module5 | 100.0 | 1456 | 0 | 103 | 1573 | 12.5% | 3.00 | 12.5 | 0.010 |

Testing using fault injection is done many times in a certain time. The testing aims to know will the system run stable without fault, will the system have a correct result / output and can the system detects the error and recover the module which is an error. The testing is done by giving the fault injection in a periodic time every one second to random module. Fault injection is given to module 1 to 5 and also fault injection is given to two modules at the same time.

Image 6 is the structure of each module. Each module contains extension hamming code calculation to decode and encode the data. The input is the data that is to be encoded, the output will be encoded data and decoded data of hamming code. By that structure of each module we can know how the voter and how the error detector will work. Can the voter compare the data from the output of each module to get valid result and can the error detector detect module which is an error.

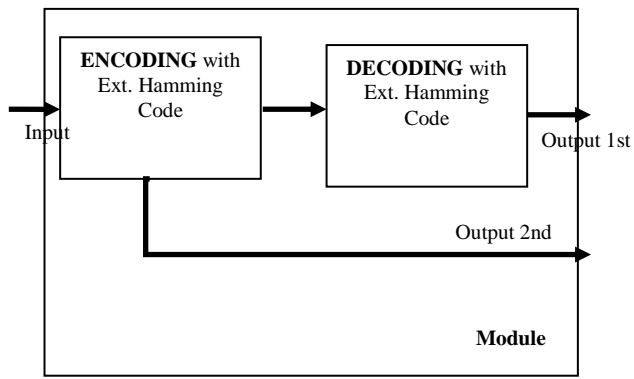

Image 6. The structure of each module

In image 7 showed there are three important cycles. First is shown that the FMR system is injected the blank module, second when the FMR consider having corrupt module we send the data to system, then the system will send to each module to do hamming code calculation. Each module will encode and decode, and give the output to the voter. The voter now has a data and sends the data to output. Third, the fault tolerance system detected the error and then the system will do DPR to mitigate the error module. From those cycles we showed in the image 7 that the system was injected by blank module to module **3, 4 and 5,** because we injected blank module, now those modules were corrupted. To know whether the system can do the operation if there are error modules, we send the data (**1010**) to the system immediately before the system do recovery. Each module in the system is encoding and decoding, and sends the data to voter. The data from voter is sent to a computer to be evaluated, it showed **00100101** for encoding and **1010** for decoding**,** even we introduce error in the first bit of encoding data the system still can correct the data which is an error by hamming code. After processing the data the system detects the error in a module and continues to recovery error modules. It showed modules **3, 4 and 5** were recovered.

Those cycles are done for more than one hour and the system did DPR more than 3600 times, the result we got that the system is stable, can correct the error module and give output with correct data as well.

| Time | Fault Injection / Blank | Recovery | Decoding Data | Encoding Data | DPR Speed |
|---|---|---|---|---|---|
| 04:25:21.511 | | | | | 141.56 |
| 04:25:21.368 | | 3 | | | |
| 04:25:21.368 | | | 1010 | 00100101 | |
| 04:25:21.367 | | | | | 147.11 |
| 04:25:21.216 | 3 | | | | |
| 04:25:20.728 | | | | | 252.72 |
| 04:25:20.467 | | | 1010 | 00100101 | |
| 04:25:20.467 | | 5 | | | |
| 04:25:20.467 | | | | | 261.49 |
| 04:25:20.206 | 5 | | | | |
| 04:25:20.173 | | | 1010 | 00100101 | |
| 04:25:20.172 | | | | | 252.70 |
| 04:25:19.921 | | 5 | | | |
| 04:25:19.921 | | | | | 224.95 |
| 04:25:19.690 | | 4 | | | |
| 04:25:19.690 | | | 1010 | 00100101 | |
| 04:25:19.690 | | | | | 261.49 |
| 04:25:19.424 | 5 | | | | |
| 04:25:19.424 | | | | | 233.10 |
| 04:25:19.188 | 4 | | | | |

Image 7. Testing result of fault injection to FMR

## V. Conclusion

Five Modular Redundancy (FMR) can be implemented successfully, when the error is happening in two modules, the system still working properly. With mitigation technique such error in module can be detected and corrected so that the fault tolerance system will try to keep away from degradation in each module. Using DPR, Five Modular Redundancy with mitigation technique can be implemented successfully. The speed to do recovering depends on the size of the Bitsream. From *Xilinx Power Estimator*, compare to TMR, FMR power consumption is slightly more, this because we add two modules in the FMR.

## VI. Future Study

Future studies will add various functions to each module, for example by putting the module with Microblaze processor to do some tasks. Implement the fault tolerance into a real application for On Board Computer of micro satellite. Testing and analyze the behavior of FMR in a real application. Scope for future study will be about implementing the FMR into a real application based on FPGA with DPR technology.

### Acknowledgment

This project is supported by Satellite Centre - LAPAN and Doctorate Program in Computer Science, Department of Computer Science and Electronics, Faculty of Mathematics and Natural Sciences, Gadjah Mada University. We would like to acknowledge for their support in this project.